\documentclass[%
 aip,
apl,%
 amsmath,amssymb,
 reprint,%
]{revtex4-1}

\usepackage{graphicx}
\usepackage{dcolumn}
\usepackage{bm}

\begin{document}

\preprint{AIP/123-QED}

\title[Title]{An equation for the quench propagation velocity valid for high field magnet use of REBCO coated conductors}

\author{M. Bonura}\affiliation{Department of Applied Physics (GAP) and Department of Quantum Matter Physics (DQMP), University of Geneva, quai Ernest Ansermet 24, CH-1211 Geneva, Switzerland.}
\author{C. Senatore}\affiliation{Department of Applied Physics (GAP) and Department of Quantum Matter Physics (DQMP), University of Geneva, quai Ernest Ansermet 24, CH-1211 Geneva, Switzerland.}


\begin{abstract}
Based on a study of the thermophysical properties, we derived a practical formula for the normal zone propagation velocity appropriate for REBa$_2$Cu$_3$O$_{7-x}$ coated conductors in high magnetic fields. An analytical expression to evaluate the current sharing temperature as a function of the operating conditions is also proposed. The presented study has allowed us to account for experimental results not fully understood in the framework of the models widely used in the literature. In particular, we provided a fundamental understanding of the experimental evidence that the normal zone propagation velocity in REBa$_2$Cu$_3$O$_{7-x}$ coated conductors can be mainly determined by the operating current, regardless of the applied field and temperature.
\end{abstract}

\keywords{Quench propagation velocity, coated conductors, HTS, REBCO, NZPV, CCs}
\maketitle

Second generation high temperature superconductors (2G HTS), i.e. REBa$_2$Cu$_3$O$_{7-x}$ (REBCO) coated conductors (CCs), are paving the way for the development of superconducting magnets that exceed the limits of LTS based technologies.\cite{Weijers,Sangwon Yoon,Senatore1,Rossi} HTS are not subjected to stability issues as LTS, and can in principle be operated without any protection.\cite{Iwasa} However, this is usually not the case for HTS magnets, whose protection against quench-induced damages is motivated by their high cost. Different strategies are available. Active protection by distributed quench heaters is the one chosen for the 32~T magnet under construction at the NHMFL.\cite{Weijers,Markiewicz} Coils made using the no-insulation winding technique revealed to be self protecting, since the current can automatically bypass the normal spot.\cite{Hahn,Sangwon Yoon}

The normal zone propagation velocity ($NZPV$), i.e. the velocity of the superconducting/normal boundary during a quench, plays a key role in the quench process. The longitudinal component of the $NZPV$ assumes values two/three orders of magnitude smaller in 2G HTS than in LTS.\cite{Iwasa} This makes the quench detection a critical issue for HTS magnets.\cite{Marchevsky} The propagation of the perturbation along the conductor depends on the properties of the conductor itself whilst the transverse component of the $NZPV$ is also determined by any other materials possibly present in the winding. Low values for the longitudinal $NZPV$ make the quench propagation in the transverse direction more relevant for HTS than for LTS. Properties of CCs are evolving very rapidly because of the intensive research carried out by industry and University research centers. Hence, \textit{reliable} and \textit{practical} tools to determine the $NZPV$ as a function of the conductor properties are highly desirable. This is the main motivation of this letter, which proposes a new analytical procedure to evaluate the longitudinal $NZPV$ in CCs with an accuracy higher than that associated with formulas commonly used in the literature.\cite{Iwasa,Wilson} The results of our study have been compared with experimental data obtained in a commercial CC manufactured by SuperPower.\cite{Nugteren}

The $NZPV$ can be evaluated following different approaches. It can be measured experimentally in conditions as close as possible to those realized in a winding. However, measurements are complex and time consuming. \cite{Nugteren} Numerical simulations are another way to assess the $NZPV$, particularly useful in case of systems with complex design.\cite{Haro}
The most practical way to evaluate the $NZPV$ is to use the formulas that result from the solution of the heat equation describing the quench process.\cite{Iwasa,Wilson,Dresner1} A good agreement between experimental and calculated values has been demonstrated in case of LTS. Nevertheless, the validity of the approximations made when solving the equations may fail for HTS. \cite{Iwasa,Wilson} In spite of this, scientists and magnet designers continue using confidently the analytical approach to reach an understanding of quenches in HTS, due to the difficulties encountered when evaluating the $NZPV$ experimentally or numerically. Based on the experimental study of the thermal conduction properties $\kappa(T,B)$, in the following we derive a practical equation for the $NZPV$ appropriate for 2G HTS in presence of high magnetic fields.

The differential equation describing the adiabatic quench process in a superconductor is:\cite{Iwasa}
\begin{equation}\label{Eq_Heat}
      C(T)\frac{\partial T}{\partial t}=\nabla\cdot[\kappa(T)\nabla T]+g_j(T) .
\end{equation}
In Eq.~(\ref{Eq_Heat}), the left-hand side represents the variation in the internal energy density of the conductor, $C$ being the volumetric specific heat. On the right-hand side, the first term describes the thermal conduction in the conductor; the second term is the Joule heating, whose explicit form in case of composite superconductors is $g_j(T)=\rho_m(T)J_m(T)J$, where $\rho_m$ is the matrix electrical resistivity, $J_m$ and $J$ the current density in the matrix and in the composite, respectively.\cite{Bellis}
Whetstone and Roos derived an analytical expression for the $NZPV$ in adiabatic conditions, assuming that the normal-superconducting boundary during a quench can be represented by a translating coordinate system moving at $NZPV$.\cite{Whetstone} The formula was successively modified by Bellis and Iwasa in order to take into account the effects due to the current sharing between superconductor and metal matrix.\cite{Bellis} This adaptation is particularly important in the case of HTS because the temperature range over which the current sharing occurs is much larger than in LTS. \cite{Bellis} The deduced expression is:
\begin{eqnarray}\label{Eq_W-R}
NZPV\approx J\bigg[\frac{1}{\rho_n(T_t)\kappa_n(T_t)}\bigg(C_n(T_t)-\frac{1}{\kappa_n(T_t)}\frac{d\kappa_n}{dT}\bigg|_{T=T_t} \nonumber\\
\times\int_{T_{Op}}^{T_t}C_S(T)dT\bigg)\int_{T_{Op}}^{T_t}C_S(T)dT\bigg]^{-1/2}  .
\end{eqnarray}
The subscripts \textit{n} and \textit{s} refer to the normal and superconducting state, respectively. However, in REBCO CCs one can consider that $C_n\approx C_s$ and $\kappa_n \approx \kappa_s$, because variations in the overall specific heat and thermal conductivity due to the transition from the superconducting to the normal state are negligible. \cite{Bonura1,Grabovickic} The transition temperature, $T_t$, has been introduced in place of $T_C$ in Eq.~(\ref{Eq_W-R}) in order to define an effective superconducting/normal boundary during a quench when current sharing effects are important. In LTS, $T_t$ is normally considered to be the average value between the temperature at which the current sharing starts ($T_{CS}$), and the critical temperature ($T_C$), i.e. $T_t\equiv (T_{CS}+T_C)/2$. In HTS,  $T_t$ is more properly evaluated as the temperature at which the heat generation term in Eq.~(\ref{Eq_Heat}) assumes its average value in the current sharing temperature range, i.e. when
\begin{equation}\label{Eq_g_j}
      g_j(T_t)=\bigg[\int_{T_{CS}}^{T_C} g_j(T)dT\bigg]/(T_C-T_{CS})  .
\end{equation}
Neglecting the temperature dependence of the material properties, Eq.~(\ref{Eq_W-R}) may be simplified to
\begin{equation}\label{Eq_Iwasa}
      NZPV\approx \frac{J}{C}\bigg[\frac{\rho\kappa}{(T_t-T_{Op})}\bigg]^{1/2}  .
\end{equation}
Different approaches have been proposed to evaluate the $NZPV$ by Eq.~(\ref{Eq_Iwasa}). Iwasa has shown that for $(T_t-T_{Op})/T_{Op}\ll 1$, $C$, $\rho$, and $\kappa$ can be conveniently evaluated at $\tilde{T}\equiv(T_{op}+T_t)/2$, i.e. at the mean value between the operating and transition temperatures.\cite{Iwasa} On the other hand, Wilson has proposed to evaluate $\rho$, and $\kappa$ at $T=T_t$, using for $C$ the average value in the range $T_{Op}-T_C$.\cite{Wilson} The two procedures lead to similar values for the $NZPV$ in LTS and are in general not applicable to HTS.
Nevertheless, their use has been extended in the practice also to HTS,\cite{Iwasa,Grabovickic,Nugteren} because of the complications encountered when solving the more general Eq.~(\ref{Eq_W-R}). In particular, the calculation of the $NZPV$ from Eq.~(\ref{Eq_W-R}) is hindered by the need of details about the $\kappa(T)$ curve of the conductor at the conditions realized in a magnet.
It is worth mentioning that, in order to bypass these difficulties, an approximation of Eq.~(\ref{Eq_W-R}) alternative to Eq.~(\ref{Eq_Iwasa}) has been derived under the less stringent assumption that only the $T$ dependence of the electrical resistivity is negligible.\cite{Zhao,Joshi,Ishiyama} However, this hypothesis is correct only for LTS because their transition temperature typically falls in the low-temperature region where $\rho\approx\rho_{Res}$, the residual electrical resistivity.

\begin{figure}[t,b]
\centering \includegraphics[width=6 cm]{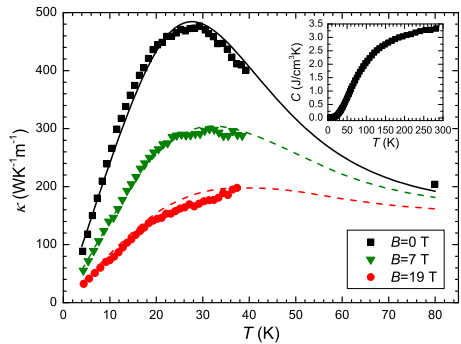} \caption{\label{Fig1}
Temperature dependence of the longitudinal thermal conductivity at different fields for the CC from SuperPower (product ID: SC S4050). The field has been applied parallel to the tape surface, perpendicularly to the thermal-flow direction. Inset: specific heat data as measured on the CC.}
\end{figure}
Recently, we reported on the thermal conduction properties of REBCO CCs from different manufacturers in magnetic fields up to 19~T.\cite{Bonura1}
In Figure~1, the experimental $\kappa(T)$ curves of the CC from SuperPower are shown as measured at $B=0$, 7 and 19~T. The solid line is the best-fit curve obtained by fitting the experimental data at $B=0$ considering that $\kappa\approx \kappa_{Cu}\cdot s_{Cu}$.\cite{Bonura1} $\kappa_{Cu}$ is the thermal conductivity of the copper, whose dependence on the $RRR$ is described in Ref.\cite{Simon}, and $s_{Cu}$ is the conductor's cross-section fraction occupied by the stabilizer.
Dashed lines, associated with in-field data, have been calculated in the framework of the Wiedemann-Franz law supposing that $\kappa(T,B)\approx [\rho(T,0)/\rho(T,B)] \kappa(T,0)$ and using magnetoresistance data measured on Cu specimens extracted from the CC, as described in Ref.\cite{Bonura1}. At low $T$, electron-defect scattering processes dominate the heat propagation. Thus, the effect of the magnetic field is in some way analogous to that of disorder in the system: both reduce the electron mean free path and, consequently, $\kappa$. On increasing $T$, the field-induced effects on $\kappa$ become less important and the $\kappa(T)$ curves associated with different $B$ values approach each other. This is a consequence of the fact that electron-phonon scattering events start to be more relevant than electron-defect ones in determining the heat conduction for $T\gtrsim50$~K.

Data reported in Figure~1 exhibit typical features of $\kappa(T,B)$ curves of REBCO CCs produced by different manufacturers,\cite{Bonura1} and provide us the necessary understanding to formulate a practical expression for the $NZPV$ suitable for 2G HTS. $T_t$ assumes values higher than $\approx 45$~K in HTS. In this range of temperatures, the derivative of $\kappa(T)$ is strongly reduced on increasing the field, as implied by Figure~1. It follows that in Eq.~(\ref{Eq_W-R}) the term $\frac{1}{\kappa_n(T_t)}\frac{d\kappa_n}{dT}|_{T=T_t}\times\int_{T_{Op}}^{T_t}C_S(T)dT$ becomes negligible with respect to $C_n(T_t)$ in case of operation at intense fields, and Eq.~(\ref{Eq_W-R}) can be approximated by:
\begin{eqnarray}\label{Eq_New}
NZPV\approx J \bigg[\frac{\rho(T_t)\kappa(T_t)}{C(T_t)\int_{T_{Op}}^{T_t}C(T)dT}\bigg]^{1/2}\, .
\end{eqnarray}
In Figure~2, the relative error made when using Eq.~(\ref{Eq_New}) in place of Eq.~(\ref{Eq_W-R}) is plotted as a function of $T_t$, for different operating conditions ($T_{Op},B$), as determined for the SuperPower tape. The error decreases on augmenting $B$ and is always smaller than $7\%$ at 19~T. Results very similar to those reported in Figure~2 have been obtained for the CCs from other manufacturers investigated in Ref.~\cite{Bonura1}. Indeed, the validity of the approximation that leads to Eq.~(\ref{Eq_New}) relies on the $\kappa(T,B)$ properties of copper, which gives the predominant contribution to the overall thermal conductivity of the tape. Thus, Eq.~(\ref{Eq_New}) can be generally used to study quench processes in Cu-stabilized 2G HTS in the presence of intense fields.
For the sake of completeness, we want to mention that Dresner published in 1994 a study in which closed formulas for the $NZPV$ are derived considering specific dependencies of the specific heat on the temperature. In the general case of an arbitrary dependence of $C$ on $T$, he proposed to solve Eq.~(\ref{Eq_Heat}) disregarding the entire term $\nabla\cdot[\kappa(T)\nabla T]$ when $T>T_C$. This corresponds to neglect not only the term $\kappa\nabla^2T$ as done (and justified) by Whetstone and Roos but also the term $\nabla\kappa\nabla T$. These assumptions lead to an expression for the $NZPV$ formally analogous to Eq.~(\ref{Eq_New}), with $T_C$ in the place of $T_t$, since the author did not consider the current sharing effect.\cite{Dresner2}
However, Dresner did not justify the hypothesis of neglecting the $T$ dependence of $\kappa$. The validity of this assumption, which leads to Eq.~(\ref{Eq_New}), has been fully demonstrated in this letter in the case of Cu-stabilized REBCO CCs submitted to intense fields.
\begin{figure}[t,b]
\centering \includegraphics[width=6 cm]{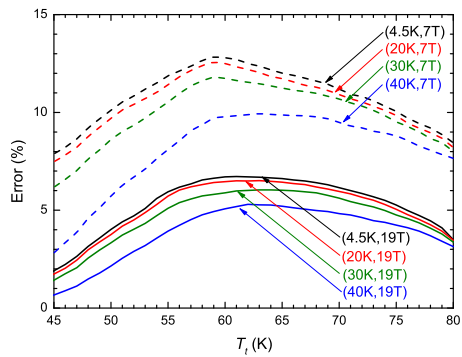} \caption{\label{Fig2}
Relative error when using Eq.~(\ref{Eq_New}) in the place of the more general Eq.~(\ref{Eq_W-R}) for different operating conditions ($T_{Op},B$), as determined for the CC from SuperPower.}
\end{figure}

Eq.~(\ref{Eq_New}) can be further simplified considering that in the framework of the Wiedemann-Franz law $\rho(T_t)\kappa(T_t)\approx LT_t$. This reduces the parameters needed to perform the calculation to: $L$, $C$, and $T_t$ (apart from $J$). $L$ values of REBCO CCs are available in the literature\cite{Bonura1}. It has been shown that $L$ does not depend noticeably on $B$.\cite{Bonura1} The specific heat of CCs can be calculated from data of the component materials, considering that $C(T)=\sum v_iC_i(T)$, $v_i$ being the volume fraction occupied by the $i^{th}$ component. It is expected that the predominant contributions come from the substrate and the stabilizer, because of the large $v_i$ values. However, we have investigated experimentally the $C(T)$ curve using a Quantum Design PPMS, in order to get more precise results. Data relative to the tape from SuperPower are shown in the inset of Figure~1. Details on the critical current surface of the CC are needed to determine $T_t$. Recently, it has been shown that the $T$ dependence of $J_C$ of CCs from different manufacturers can be described over a broad range of temperatures and fields by an exponential law, $J_{C}(T,B)=J_{C}(T=0,B)e^{-T/T^*}$. Deviations from this behavior are observed at temperatures $\gtrsim 50$~K.\cite{Senatore_sub} The exponential dependence of $J_C$ is connected with the presence of defects generating weak isotropic pinning and $T^*$ is the characteristic pinning energy at these defects.\cite{Puig,Gutierrez} Other dominant pinning mechanisms can lead to different $J_C(T,B)$ characteristics.\cite{Strickland}
The $T^*$ values associated with tapes from different manufacturers, for different orientations between the field and the tape surface, are reported in Ref.\cite{Senatore_sub}. From the expression for $J_{C}(T,B)$ one can easily deduce the following formula that relates $T_{CS}$ to parameters directly chosen by the magnet designer, namely the operating conditions and the current margin
\begin{equation}\label{Eq_TCS}
      T_{CS}\approx T_{Op}-T^*ln\frac{I_{Op}}{I_C(B_{Op},T_{Op})}  \, .
\end{equation}
\begin{figure}[t,b]
\centering \includegraphics[width=6 cm]{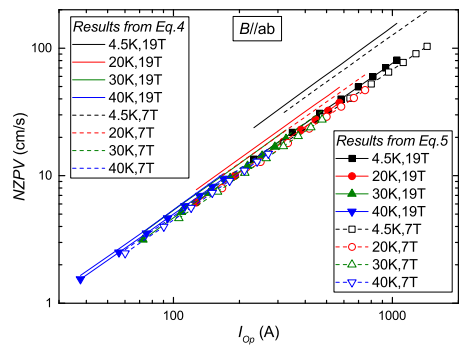} \caption{\label{Fig3}
Dependence of the $NZPV$ on the operating current at different $T_{Op}$ and $B$, deduced from Eq.~(\ref{Eq_New}) (lines and symbols) and from Eq.~(\ref{Eq_Iwasa}) following the procedure described by Iwasa\cite{Iwasa} (lines without symbols).}
\end{figure}
$T_t$ can be evaluated from Eq.~(\ref{Eq_g_j}), using $T_{CS}$ values from Eq.~(\ref{Eq_TCS}) and the expression for $J_C(T,B)$ reported in Ref.~\cite{Senatore_sub}. We have also verified that the approximated formula $T_t\equiv (T_{CS}+T_C)/2$ leads to a good estimation for $T_t$, with differences within $\approx 5\%$, when $T\gtrsim 20$~K. At 4.5~K, the discrepancies increase up to $\approx 10\%$.

\begin{table}[t,b]
\caption{\label{Tab1}Critical Current Densities of the investigated CC}
\begin{ruledtabular}
\begin{tabular}{ll}
$I_C$(4.5K,7T)$\approx1600$~A & $I_C$(4.5K,19T)$\approx1160$~A \\
$I_C$(20K,7T)$\approx830$~A & $I_C$(20K,19T)$\approx640$~A \\
$I_C$(30K,7T)$\approx530$~A & $I_C$(30K,19T)$\approx360$~A \\
$I_C$(40K,7T)$\approx300$~A & $I_C$(40K,19T)$\approx190$~A \\
\end{tabular}
\end{ruledtabular}
\end{table}

In Figure~3, we report the $NZPV$ as determined from Eq.~(\ref{Eq_New}) considering that $\rho(T_t)\kappa(T_t)\approx L T_t$, using $L$ values reported in Ref.~\cite{Bonura1}, $T_{t}$ calculated using the definition given in Eq.~(\ref{Eq_g_j}), and experimental $C(T)$ data shown in the inset of Figure~1. $I_{Op}$ has been varied in the range 0.2-0.9 $I_C$, using for $I_C(T,B)$ the values reported in Table~1. These have been measured on CCs extracted from the same batch of the sample used for the thermal conduction studies, with the field applied parallel to the wide surface of the tape.
Figure~3 shows that the $NZPV$ is mainly determined by the operating current. Points associated with different temperatures and fields approximatively reconstruct a single line in a log-log plot, defining a power-law dependence of the $NZPV$ on $I_{Op}$. This behavior is unexpected if compared to what observed in LTS. Indeed, both in NbTi and Nb$_3$Sn a clear dependence of the $NZPV$ on $B$ is found.\cite{Zhao}
The contrast between the result shown in Figure~3 and what is observed in LTS is certainly related with the different $I_C(T,B)$ characteristics of the materials. $NZPV$ values calculated from Eq. (4) following the procedure described by Iwasa\cite{Iwasa} are shown in Figure 3 as lines without symbols.
Data associated with different operating conditions do not lie all on a same straight line in a log-log plot. Discrepancies between results from Eq.~(\ref{Eq_Iwasa}) and Eq.~(\ref{Eq_New}) become more evident, both qualitatively and quantitatively, on decreasing the operating temperature. This is worth to underline in view of applications of CCs in very high field magnets. Our results about the dependence of the $NZPV$ on $I_{Op}$ at different operating conditions are confirmed by the experimental $NZPV$ studies performed on a SuperPower tape extracted from another batch with respect to ours.\cite{Nugteren} The experimental confirmation strengthens the validity of the analytical procedure proposed in this letter to determine the $NZPV$.
When comparing quantitatively experimental data with theoretical expectations, one has to take into account the so-called \textit{minimum propagation current} ($I_{mp}$) i.e. the operating current below which there is no quench triggering even for pulses with an energy exceeding the stability margin.\cite{Grabovickic,Nugteren} $I_{mp}$ values reported in the literature for CCs are in the range $10-30$~A.\cite{Grabovickic} The effect of $I_{mp}$ on the measured $NZPV$ can be neglected when $I_{Op}\gg I_{mp}$. Samples investigated by us and in Ref.\cite{Nugteren} present slightly different $I_C$ characteristics.
Nevertheless, we verified that for $I_{Op}\gtrsim 100$~A, which is much larger than expected $I_{mp}$, discrepancies between values shown in Figure~3 and data from Ref.\cite{Nugteren} are below 25\%.

Results from Eq.~(\ref{Eq_New}), combined with longitudinal and transverse $\kappa$ data, allow calculating the transverse $NZPV$.\cite{Iwasa,Wilson} In Ref.\cite{Bonura2} we have reported experimental values for the square root of the ratio between the transverse and longitudinal components of $\kappa$ for various CCs. Typical values are of the order of 0.1. These data provide lower limits for the anisotropy of the $NZPV$ in a winding, since the contact thermal resistance or the presence of other materials, which could reduce the overall transverse $\kappa$, have not been considered.

In summary, an approximated equation has been derived for the longitudinal $NZPV$, Eq.~(\ref{Eq_New}), particularly suitable for 2G HTS in intense magnetic fields. An analytical expression to evaluate the current sharing temperature as a function of the operating conditions, Eq.~(\ref{Eq_TCS}), has also been proposed. The presented study has allowed us to take into account experimental results not fully understood in the framework of models widely used in the literature.

Financial support was provided by the SNSF (Grants No. PP00P2$\_$144673 and No.51NF40-144613) and by FP7 EuCARD-2. EuCARD-2 is cofounded by the partners and the European Commission under Capacities 7th Framework Programme, Grant Agreement 312453. The authors thank Piotr Komorowski and Christian Barth for useful discussions.

\end{document}